\documentclass[twocolumn]{article}

\newif\ifbandfig
\bandfigtrue
\usepackage{times}

\usepackage{amsmath}
\usepackage{color}
\newif\ifdraft
\usepackage{graphicx}
\usepackage{lineno}
\usepackage[affil-it]{authblk}
\usepackage[colorlinks,citecolor=blue,breaklinks=true]{hyperref}

\newcommand{\etc}[0]         {etc.}
\newcommand{\eg}[0]          {e.g.}
\newcommand{\ie}[0]          {i.e.}
\newcommand{\cf}[0]          {\textit{cf.}}
\providecommand{\etal}[0]    {\textit{et al.}}

\newcommand\bx               {\boldsymbol x}
\newcommand\bV               {\boldsymbol{V}}
\newcommand\bv               {\boldsymbol{v}}
\newcommand\vopar            {\boldsymbol{M}}
\newcommand\opar             {M}
\newcommand{\entrop}         {{\mathbf H}}
\newcommand\IOP              {\mathcal I}
\newcommand\Ipw              {\IOP_{pw}}
\newcommand\IpwLT            {\IOP^{LT}_{pw}}
\providecommand{\e}[1]       {\ensuremath{\times 10^{#1}}}
\renewcommand\phi            {\varphi}
\renewcommand\epsilon        {\varepsilon}

\newcommand{\eqnRef}[1]{Eqn.~\ref{eqn:#1}}
\newcommand{\figRef}[1]{Fig.~\ref{fig:#1}}

\title{Anomalous behaviour of mutual information in finite flocks}

\author{L. Barnett} \affil{Sackler Centre for Consciousness Science, Department of Informatics, University of Sussex, Brighton, U.K.}
\author{J. Brown \thanks{Email: \texttt{jbrown@csu.edu.au}; Corresponding author}} \affil{School of Computing \& Mathematics, Charles Sturt University, Bathurst, NSW, Australia}
\author{T. Bossomaier} \affil{Centre for Research in Complex Systems, Charles Sturt University, Bathurst, NSW, Australia}

\begin{document}

\maketitle

\begin{abstract}
The existing consensus is that flocks are
poised at criticality~\cite{vanni11,mora11}, entailing long
correlation lengths and a maximal value of Shannon  mutual
information~\cite{shannon48} in
the large-system limit.  We show, by contrast, that for \emph{finite} flocks in the long observation time limit,
mutual information may not only fail to peak at
criticality---as observed for other critical
systems~\cite{matsuda96,lau13,barnett13:prl,langton90,harre09:epl,erb04}---but
also diverge as noise tends to zero. This result carries implications
for other finite-size, out-of-equilibrium systems, where observation
times may vary widely compared to time scales of internal system
dynamics; thus it may  not be assumed that mutual
information locates the phase transition.
\end{abstract}

\section{Introduction} \label{sec:intro}

From the 40,000 strong murmurations of starlings to traffic jams,
flocking occurs in many animal species, as well as many domains of
human society. Recent developments in video pattern recognition and
GPS technology have greatly increased our understanding of animal
systems, such as fish~\cite{calovi15,rosenthal15},
pigeons~\cite{nagy13}, starlings~\cite{cavagna10:starlings},
midges~\cite{attanasi14:midges} and sheep~\cite{ginelli15}. Flocks
offer energy efficiency, reduced navigational effort and increased
resilience to predation.  For biological, finite-size flocks, however, it is
under-appreciated that macroscopic statistics depend essentially on
observation time scales.

Understanding of flocking dynamics owes much
to abstract models, such as  the standard Vicsek
model (SVM)~\cite{vicsek95} which, at large system size, exhibits phase transition-like behaviour
at a critical noise value. In other  systems studied to date, such as the Ising spin
model~\cite{matsuda96,lau13,barnett13:prl}, cellular
automata~\cite{langton90} and financial systems~\cite{harre09:epl},
mutual information (MI) is the gold-standard marker of order-disorder
($2$nd order) phase transitions in equilibrium statistical mechanics:
in the thermodynamic limit it typically tends to zero in the limits of
low and high noise, peaking at criticality. Less is known,
however, of its behaviour in out-of-equilibrium and/or finite-size
systems. The SVM exemplifies an
out-of-equilibrium phase transition~\cite{odor04} between coordinated
behaviour and random diffusion~\cite{vicsek95}, thought to be in its
own universality class~\cite{baglietto08:fss}. The
  thermodynamic limit of large system size has been studied by Toner
  and Tu, both at the phase transition~\cite{toner95} and the
  low noise, single flock, limit~\cite{toner98:pre}. In the limit
  symmetry is broken, leaving Goldstone modes, and thus large, long-range
  density fluctuations in two dimensions. In higher dimensions, the
  situation is more complicated.

  Here, by exploiting an
approximate isometry of the SVM, we obtain a novel closed-form
dimensional reduction of the neighbour-pair MI between particle
headings on the basis that, in a finite-size system at long
observation times, rotational symmetry is never broken. This reveals a
hitherto unnoticed behaviour of MI in such systems: \emph{absence of a
  peak at the phase transition, and divergence at low noise}, contrary
to behaviour in the Ising model and other complex systems~\cite{erb04}.

\section{The standard Vicsek model}

The SVM comprises a set of $N$ point particles (labelled $i = 1,\ldots,N$) moving on a plane of linear extent $L$ with periodic boundary conditions. Each particle moves with constant speed $v$, and interacts only with neighbouring particles within a fixed radius $r$, which we take to be $1$. We denote the position of the $i$th particle by $\bx_i(t)$ and its velocity vector by $\bv_i(t) = (v\cos\theta_i(t), v\sin\theta_i(t))$, where $\theta_i(t)$ is its \emph{heading}~\footnote{We consider headings as \emph{circular} variables defined on $(-\pi,\pi]$ with arithmetic modulo $2\pi$.}. Let $\nu_i(t) \equiv \{j \,:\, |\bx_j(t)-\bx_i(t)| < r\}$ be the index set of all particles neighbouring particle $i$ at time $t$ (including $i$ itself, so that $\nu_i(t) \ne \emptyset$). The \emph{neighbourhood-average velocity} of particle $i$ is then given by:
\begin{linenomath}
\begin{equation}
	\bar\bv_i(t) = \frac{1}{|\nu_i(t)|}\sum_{j \in \nu_i(t)} \bv_j(t)\,,
\end{equation}
\end{linenomath}
with heading $\bar\theta_i(t)$.

Particle positions and headings are updated synchronously~\footnote{We implement a ``backward update'' scheme, where both particle positions and velocities for time $t+\Delta t$ are updated on the basis of particle velocities at time $t$, as opposed to the ``forward update'' scheme which updates particle positions for time $t+\Delta t$ using the already updated velocity at $t+\Delta t$.} at discrete time intervals $\Delta t = 1$ according to
\begin{linenomath}
\begin{align}
	\bx_i(t + \Delta t) = \bx_i(t) + \bv_i(t)\Delta t \label{eqn:OVAposUpdate}\,, \\
	\theta_i(t+\Delta t) = \bar\theta_i(t) + \omega_i(t) \label{eqn:OVAvelUpdate}\,,
\end{align}
\end{linenomath}
respectively, where $\omega_i(t)$ is a thermal fluctuation (white noise) uniform on the interval $[-\eta/2, \eta/2]$ with intensity $\eta \in (0,2\pi]$.

Note that, since a particle travels a distance $v$ in a single time increment $\Delta t = 1$, the SVM only approximates continuity in space and time in case $v \ll 1$ (the model is thus arguably unrealistic as a model for real-world flocking if particle velocities are large).

\subsection{The SVM ensemble}

We consider the SVM as a \emph{statistical ensemble} of finite size $N$, parametrised by the velocity $v$, particle density $\rho \equiv L^2/N$ and noise intensity $\eta$. For simplicity, particle density is fixed at $\rho = 0.25$ throughout, and noise intensity $\eta$ is taken as a control parameter. We suppose that the ensemble is relaxed into a steady state, and use capitals $\bV_i, \Theta_i$, \etc, to indicate corresponding quantities sampled from the steady-state ensemble. Angle brackets $\langle\cdots\rangle$ denote ensemble averages. In the limit $v \to 0$, the model is equivalent to an \emph{XY} model, where particles do not move~\footnote{Note that the limiting behaviour of the model as $v \to 0$ must be considered as distinct from models with $v = 0$~\cite{baglietto09:cpc}.}, while in the limit $v \to \infty$ particles become fully mixed between updates~\cite{vicsek95}.

The full order parameter for the SVM ensemble is the $2$D random vector
\begin{linenomath}
\begin{equation}
	\vopar = \frac1{Nv} \sum^N_{i=1}\bV_i\,, \label{eqn:orderParameterVec}
\end{equation}
\end{linenomath}
with magnitude $\opar \equiv |\vopar|$ and heading $\Phi$. We have $0 \le \opar \le 1$, with $\opar = 1$ if and only if all particles in the ensemble are aligned, and $\opar \to 0$ in the large-system limit $N \to \infty$. The ensemble variance
\begin{linenomath}
\begin{equation}
	\chi = \langle \opar^2\rangle-\langle \opar\rangle^2 \label{eqn:susc}
\end{equation}
\end{linenomath}
of the order parameter magnitude defines the susceptibility. Although phase transitions only exist formally in the thermodynamic limit, for finite systems we consider a peak in susceptibility (with respect to a control parameter) as identifying the approximate location of a phase transition.

\subsection{\label{sec:stats}Long-term vs. short-term statistics}

In estimating ensemble statistics from simulated (steady-state) dynamics, it is commonplace to invoke \emph{ergodicity} in some form: that is, the simulation is observed, and statistics collated, over a time window of length $T$, under the assumption that as $T \to \infty$ the statistic in question converges to its ensemble average value. This approach implicitly assumes that observation times are long in comparison to the internal dynamics of the system. In the case of the finite-size SVM, however, this assumption may well be violated, particularly at low noise intensities. What we see, rather, is akin to what has been termed ``continuous ergodicity-breaking''~\cite{mauro07}: over short observation times, the system is confined to a comparatively small volume of phase space. As we observe the system over increasing lengths of time, progressively larger volumes of phase space are explored. Since a finite SVM is ergodic, the system eventually explores the entire phase space. At low noise, however, observation times necessary to obtain effectively ergodic behaviour become impractically large.

Our resolution to this issue is a pragmatic one: we consider ensemble statistics as essentially observation time-dependent. \emph{Short-term} statistics are thus collated separately (with no ergodic assumptions) over ranges of observation times spanning several orders of magnitude. This affords insights into how the extent of phase space exploration affects our statistics (and also neatly side-steps the somewhat vexed issue as to whether the SVM features true ergodicity-breaking in the thermodynamic limit). In addition, to estimate the limiting ergodic behaviour of the system, below we exploit a rotational symmetry approximation to collate \emph{long-term} statistics, under the assumption that in a finite-size SVM, symmetry---like ergodicity---is never truly broken.

\subsection{Simulation details}

Simulation models were written in C++ and run on the \textsc{raijin} supercluster at the Australian National Computer Infrastructure Facility. Since the particle velocity (angle) is continuous, the mutual information, was  calculated using nearest neighbour estimators~\cite{kraskov04,gomez-herrero15}. The accuracy of the estimators was checked by: permutation testing---shuffling the source to remove any information sharing; and decimation---comparing the estimate with subsets of one tenth the number of events~\cite{Brown17}. Theoretical work on the performance of these estimators is limited and is most relevant to  smaller systems~\cite{gao16w}. The entropy estimation by nearest neighbour is computationally demanding and was carried out in situ on \textsc{raijin}. As an example of the data requirements, the interactions of particles in the large window simulation, with $N=500, T=5\e4$, at $\eta=0.1$ produced approximately $2\e9$ points for the nearest neighbours estimators, each of which requiring a \emph{k}-nn and fixed radius search.

To reduce computation times required for simulations to settle into a steady state, we employed a cooling regime, whereby simulations were started with the maximum noise ($\eta=2\pi$) case, with particles uniformly distributed over the flat torus and headings uniformly distributed on $(0, 2\pi]$. Simulations were run for an initial number $T_s$ of skip steps to allow the system to settle, followed by a data collection phase of $T$ time steps, over which MI statistics were collated. On completion, $\eta$ was decreased and another $T_s +T$ simulation steps run with the new $\eta$ value. This technique enabled reduction of $T_s$ by an order of magnitude, as compared to restarting simulations anew for each $\eta$. Appropriate settling time depends on $\eta$ ($\eta=2\pi$, for instance, requires zero settling time). We found that a satisfactory regime was to adjust $T_s$ in tiers:
\begin{linenomath}
\begin{equation}
T_s = \begin{cases}
1000 &\eta \geq 3.0\\
50000 &\eta \leq 1.0\\
20000 &\text{otherwise}
\end{cases}
\label{eqn:skipStepTiers}
\end{equation}
\end{linenomath}

\section{Neighbour-pair mutual information}

The neighbour-pair MI is defined as the ensemble statistic
\begin{linenomath}
\begin{equation}
	\Ipw \equiv \IOP(\Theta_I : \Theta_J) = \entrop(\Theta_I) + \entrop(\Theta_J) - \entrop(\Theta_I,\Theta_J) \,,\label{eqn:Ipw}
\end{equation}
\end{linenomath}
where $\entrop$ denotes \emph{differential entropy} and $(I,J)$ is uniform on the set of unique neighbour index-pairs~\footnote{$\Ipw$ is essentially the same quantity as calculated in~\cite{barnett13:prl}, although it was formulated somewhat differently there.}. Note that by particle indistinguishability, the marginal distributions of $\Theta_I$ and $\Theta_J$ are the same, so that $\entrop(\Theta_I) = \entrop(\Theta_J)$. In the short-term case, we estimate $\Ipw$ over multiple realisations of simulated SVMs. The SVMs are first relaxed/annealed to a steady state, and then headings $\theta_i(t)$ sampled over a further simulation period of $T$ time steps, where $T$ is the observation window~\cite{physrev-mi-suppmat}.

Given that (as discussed above) ergodicity remains unbroken in the long-term observation limit, near-isotropy of the SVM allows us to approximate \eqnRef{Ipw} in this case by a one-dimensional form, in which only particle heading differences $\theta_i-\theta_j$ appear. Specifically, we assume \emph{rotational symmetry}: that for any fixed angle $\phi$, the joint distribution of $(\Theta_1+\phi,\ldots,\Theta_N+\phi)$ is the same as the joint distribution of $(\Theta_1,\ldots,\Theta_N)$. We note that the SVM on the 2D torus with periodic boundary conditions is not strictly isotropic, so that this is indeed an approximation. We tested the approximation by repeating our experiments with the frame of reference of the SVM rotated randomly between updates, thus enforcing isotropy~\cite{baglietto09}. We found that in a large, but finite, SVM the isotropy assumption introduces almost negligible error (the error only being discernible near the phase transition; see the inset in \figRef{IpwLT}).

Let $p(\theta_1,\theta_2)$ be the probability density function (pdf) of $(\Theta_I,\Theta_J)$. Under assumption of rotational symmetry, we have
\begin{linenomath}
\begin{equation}
	p(\theta_1,\theta_2) = \frac1{2\pi} q(\theta_1-\theta_2)\,,
\end{equation}
\end{linenomath}
where $q(\theta)$ is the pdf of $\Theta_I-\Theta_J$. Noting that the marginal distributions of $\Theta_I$ and $\Theta_J$ are uniform on the unit circle, we obtain
\begin{linenomath}
\begin{equation}
	\entrop(\Theta_I,\Theta_J) = \log2\pi + \entrop(\Theta_I-\Theta_J)\,,
\end{equation}
\end{linenomath}
leading to the expression
\begin{linenomath}
\begin{equation}
	\IpwLT = \log2\pi - \entrop(\Theta_I-\Theta_J) \label{eqn:IpwLT}
\end{equation}
\end{linenomath}
for the approximate long-term neighbour-pair MI. Note that $\IpwLT$ vanishes precisely when $\Theta_I-\Theta_J$ is \emph{uniform}. This is the case at maximum noise, when $\Theta_I,\Theta_J$ are independent; that is, $\IpwLT$ vanishes at maximum noise, as we would expect. At very low noise, all particles nearly align so that the distribution of $\Theta_I-\Theta_J$ becomes sharply peaked. However, because $\entrop(\Theta_I-\Theta_J)$ is a \emph{differential} entropy, it will generally diverge logarithmically to $-\infty$ as the variance decreases (\eg, the differential entropy of a narrow uniform ``notch'' of width $\epsilon$ is $\log\epsilon$). Thus, in the long-term observation scenario, $\IpwLT \to +\infty$ as the noise intensity decreases to zero.

\section{Simulation Results}

\begin{figure}
\centering
\includegraphics[width=\columnwidth]{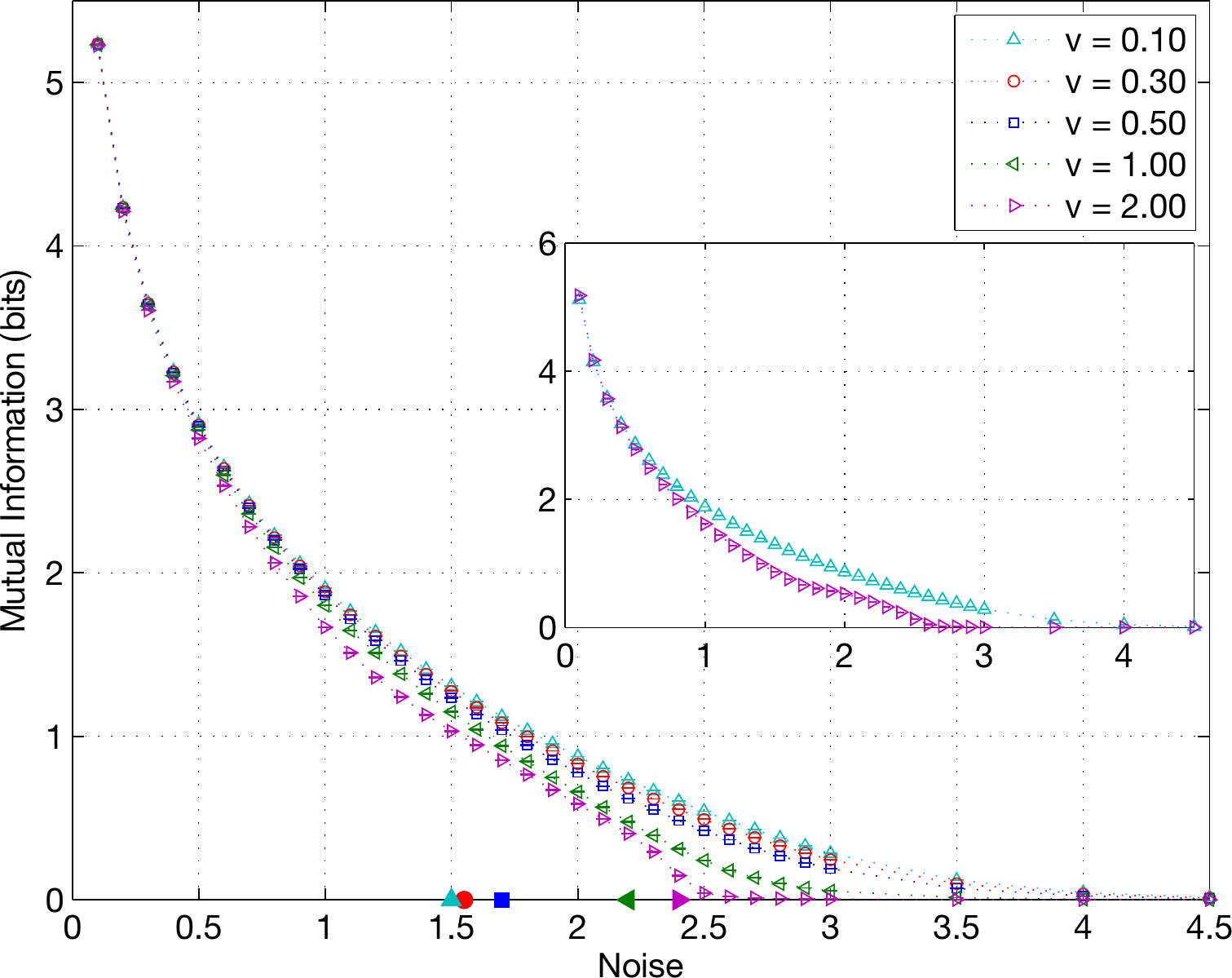}
\caption{Long-term MI $\IpwLT$ calculated according to \eqnRef{IpwLT} for a range of particle velocities. System size $N=1000$ particles, density $\rho=0.25$ and velocities $v$ as indicated. Simulation: $20$ realisations at observation time $T=500$ time steps. Error bars at $1$ s.e. (smaller than symbols) were constructed by $10$ repetitions of the experiment. $\entrop(\Theta_I-\Theta_J)$ was calculated using a $512$-bin histogram estimator (see Supplemental Material~\cite{physrev-mi-suppmat} for simulation details). Filled symbols show estimated peaks in susceptibility $\chi$. Inset: System using rotated reference frame for $v=0.10, 2.00$. } \label{fig:IpwLT}
\end{figure}

\begin{figure}
\centering
\includegraphics[width=\columnwidth]{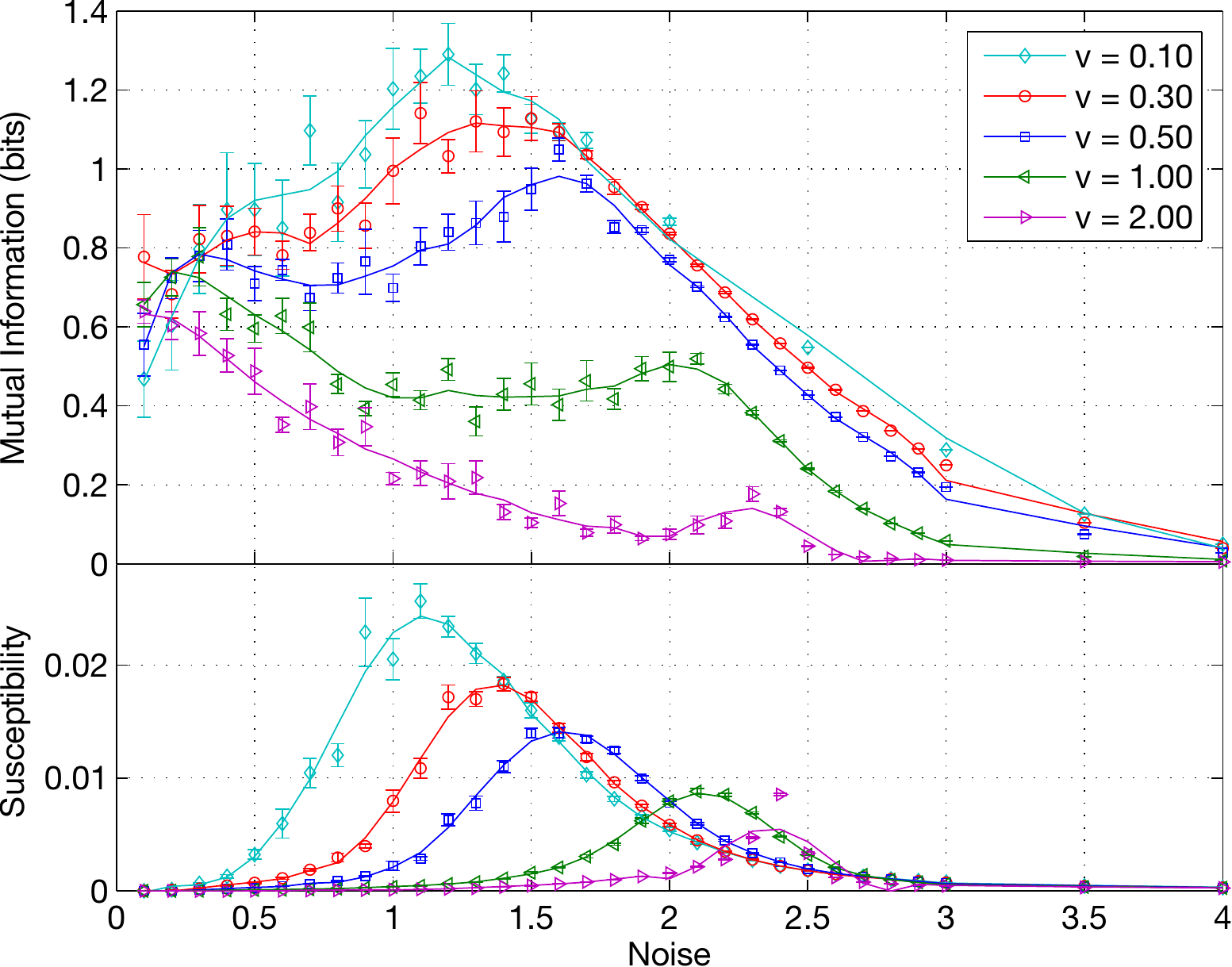}
\caption{$\Ipw$ (top plot) and susceptibility $\chi$ (bottom plot) estimated for a range of particle velocities (parameters as in \figRef{IpwLT}). $\Ipw$ was estimated according to \eqnRef{Ipw} over $T=5000$ time steps after relaxation to steady state, using a nearest-neighbour estimator~\cite{physrev-mi-suppmat}. $\chi$ was estimated over the same realisations. Error bars at $1$ s.e. again constructed by $10$ repetitions.} \label{fig:mi-2d}
\end{figure}

\begin{figure}
\centering
\includegraphics[width=\columnwidth]{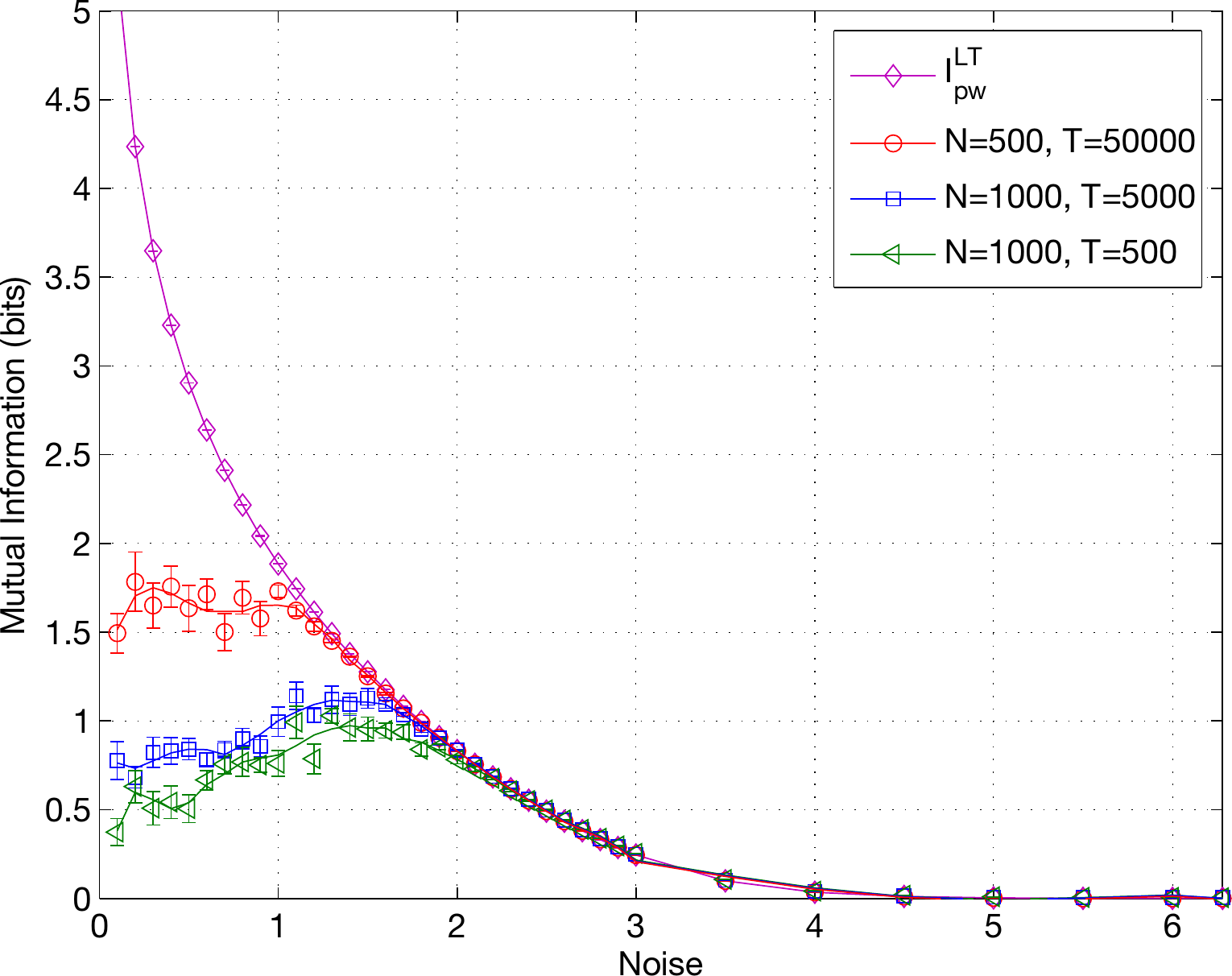}
\caption{$\Ipw$ estimates according to \eqnRef{Ipw} at fixed velocity $v = 0.30$ for a range of observation times $T$ as indicated, along with the long-term $\IpwLT$ of \eqnRef{IpwLT} as per \figRef{IpwLT}. System sizes $N$ as indicated, other simulation details as for previous figures.} \label{fig:mi-window-size}
\end{figure}

Figure \ref{fig:IpwLT} shows the long-term MI $\IpwLT$ estimated in sample according to \eqnRef{IpwLT} for a range of particle velocities. Note that there is no evidence of a peak at the phase transition.
For \emph{short observation times,} by contrast, $\Ipw$ estimated according to \eqnRef{Ipw} (\ie, with no assumption of rotational symmetry) does indeed peak at the phase transition, as reported by Wicks et al.~\cite{wicks07}; see \figRef{mi-2d}.
Some divergence at low noise is also in evidence. Figure~\ref{fig:mi-window-size} plots $\Ipw$ for a single fixed velocity at observation window size $T$ varying over two orders of magnitude, along with the long-observation-time limit $\IpwLT$. As observation time increases, the MI peak flattens and divergence at low noise increases, approaching, as predicted, the long-observation-time limit.

Thus long-term MI behaves in a distinctly different fashion to what has typically been observed in short-range MI studies of order-disorder transitions, where MI is seen to peak at the phase transition, and then tail off as noise tends to zero. Here, as the noise intensity is reduced towards zero, particles align more and more strongly, so that the distribution of $\Theta_I-\Theta_J$ becomes more and more sharply peaked, resulting in divergence of $\IpwLT$. At the same time, non-ergodicity-breaking is evidenced by a random walk-like \emph{precession} of the order parameter heading $\Phi$ around the unit circle (\cf~\figRef{bands} below).

We remark that the continuous-state nature of the SVM is central to the divergence; in a discrete-state system MI cannot diverge. Nonetheless, a similar effect is seen for discrete systems, although divergence is capped by the number of distinct states. For a discretised Vicsek system, for example, where particle headings are constrained to $n$ equispaced sectors, rotational symmetry remains unbroken (fluctuations due to noise still cause precession of $\Phi$ around the sectors) so that \eqnRef{IpwLT} still holds, with the $\log2\pi$ term replaced by $\log n$. Now $\entrop(\Theta_I-\Theta_J) \to 0$ as $\eta \to 0$, so that $\IpwLT \to \log n$.

\section{Discussion}

Since the introduction of the SVM, in which the phase transition was originally claimed to be second order, much controversy has surrounded its nature. Gregoire and Chat\'e~\cite{gregoire04} claimed on the basis of simulations that it was first order, and much discussion ensued. Seemingly small details affect the nature of the transition: type of noise statistics~\cite{chepizhko09}; forward versus backward updating (especially at high particle velocities)~\cite{nagy07}; boundary conditions associated with density bands or spin waves~\cite{aldana07}; and the cone of influence on each particle~\cite{durve16,romensky14}. In the interests of pragmatism, in this study we utilise the original SVM model (backward updating, angular noise, periodic boundary conditions and low density) over a range of velocities.

But there is an additional aspect to the phase transition beyond the
order controversy: the effect of finite size. In classical
equilibrium systems, finite size effects with $O(2)$ symmetry are known
to exhibit a random walk behaviour along the Goldstone modes at low
noise, but little is known about the active matter system considered
here~\cite{niel87,goldschmidt87}. For $d=2$, the
Mermin-Wagner theorem would lead one to suspect that there is no phase
transition, but this strictly only holds for ergodic, equilibrium
systems. Baglietto~\etal~\cite{baglietto08:finite} and Albano~\etal~\cite{albano11:rpp}, for example, discuss finite-size scaling, showing good agreement with theory for the susceptibility at the phase transition. 

\begin{figure*}
\centering
\ifbandfig
\resizebox{0.99\textwidth}{!}{\includegraphics{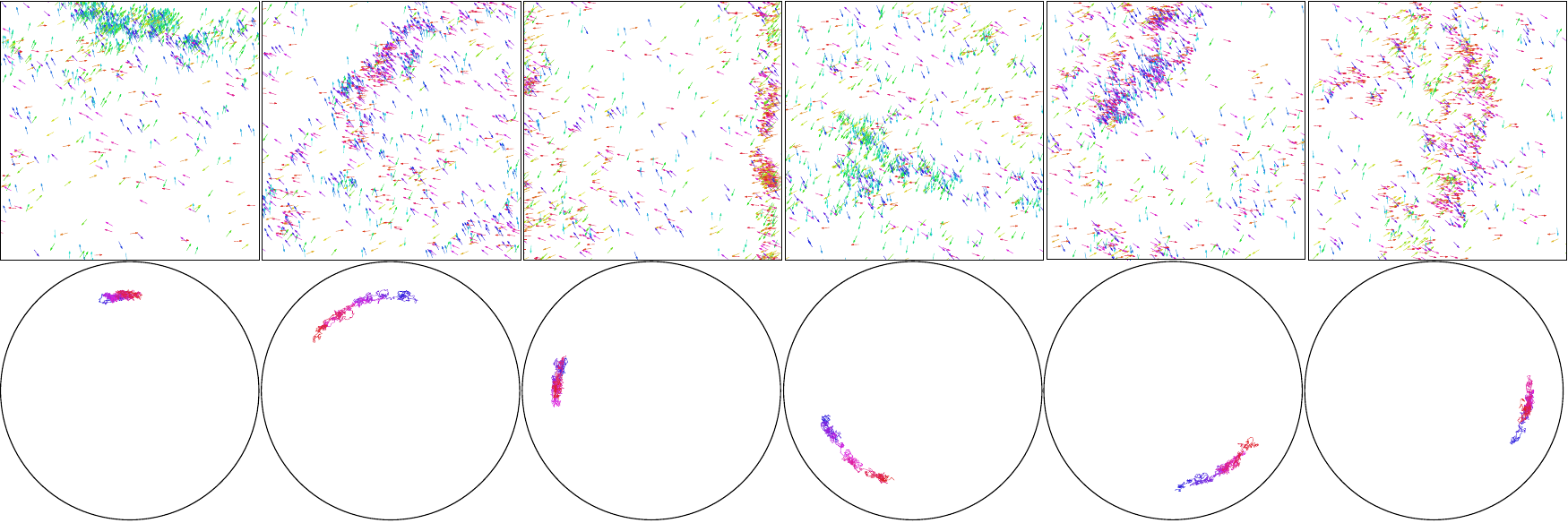}}
\fi
\caption{Snapshots from a single simulation demonstrating precession of high density bands of a flock with $N=1000$ particles at high velocity ($v=2.0$). Snapshots taken at, from left to right, $t=23\e3, 24\e3, 28\e3, 40\e3, 47\e3, 49\e3$. Top row shows the state of the flock, while the bottom row shows the two-dimensional order parameter $\vopar$---that is, mean particle velocity---for the previous $1000$ time steps going from blue ($t-1000$) to red ($t$). Distance from the centre of the circle corresponds to the order parameter magnitude $\opar = |\vopar|$. Note that, as witnessed by the first two snapshots, precession can be rapid, with only $1000$ time steps required for the band to precess $\pi/4$ radians.}
\label{fig:bands}
\end{figure*}

In the finite-size SVM, even at low (sub-critical) noise intensities, \emph{neither ergodicity nor (approximate) rotational symmetry is broken}. Rather, we see something akin to what has been described as ``continuously broken ergodicity''~\cite{mauro07}: at low noise and short observation times ergodicity is approximately broken, in the sense that the system is largely confined to small volumes of phase space. However, as observation time increases the system becomes increasingly ergodic, exploring progressively larger volumes---and ultimately the entirety---of phase space. In the finite-size SVM this manifests as a stochastic precession of the order parameter heading $\Phi$ around the unit circle (\figRef{bands}). We note too, that at (albeit physically implausible) high velocities, the SVM exhibits travelling ``bands'' of particles~\cite{nagy07}; while it might be thought that this represents true symmetry breaking, detailed simulations (\figRef{bands}) reveal that banding orientation, as well as $\Phi$, precesses.

The behaviour we see here has elements of classical thermodynamic
equilibrium systems, although it is an active matter, far-from-equilibrium system. For active matter,
the concept of equilibrium itself, is still not clearly defined~\cite{fodor16}. The continuous
symmetry implies that at even extremely low noise, the flock(s) can gradually change direction and cover the
whole of phase space as observation time tends to infinity. This movement is analogous to the
Goldstone modes of classical systems, such as the Ising model, an O(2) symmetry mode left behind
when symmetry breaking occurs. 

Goldstone modes have been discussed relative to flocking by Bialek~\etal~\cite{bialek14} and rejected as the source of information flow through the flock. On the
other hand Melfo~\cite{melfo17} claims that it is in fact the Goldstone mode which allows flock
stability over a wide range of noise (system) parameters.

Although the unexpected behaviour of the MI has been demonstrated for the SVM, it seems likely that it will apply to many finite, ergodic systems with an unbroken symmetry: for such systems, MI may vary dramatically with observational time scale, diverging in the long-term limit as thermal noise approaches zero.

\section{Acknowledgements}
We thank Mike Harr\'e, Joe Lizier and Guy Theroulaz for helpful discussions. Our attention was first drawn to the apparently anomalous behaviour of MI in the SVM by Dan Mackinlay.

The National Computing Infrastructure (NCI) facility provided computing time for the simulations under project e004, with part funding under Australian Research Council Linkage Infrastructure grant LE140100002.

Joshua Brown would like to acknowledge the support of his Ph.D. program and this work from the Australian Government Research Training Program Scholarship.

We also thank an anonymous referee for comments on the Goldstone modes.

\end{document}